\documentclass{aa}  

\usepackage{graphicx}
\usepackage{tabularx}
\usepackage{array}
\usepackage{txfonts}
\usepackage{lipsum}
\usepackage{subcaption}         
\usepackage{lscape}
\usepackage{placeins}

\begin{document} 
   \titlerunning{Water enhances the formation of refractory sulfur in interstellar ices}
   \title{Water enhances the formation of refractory sulfur in interstellar ices}
   
    \authorrunning{C. del Burgo Olivares et al.}
   \author{C. del Burgo Olivares \inst{1,2}\corrauth{ cdelburgo@cab.inta-csic.es}
          \and H. Carrascosa\inst{1}\email{hcarrascosa@cab.inta-csic.es}
          \and G. M. Mu\~noz Caro\inst{1}\email{munozcg@cab.inta-csic.es}
          \and R. Lebrón-Aguilar\inst{3}\email{rlebron@iqf.csic.es}
          \and J. E. Quintanilla-López\inst{3}\email{je.quintanilla@iqf.csic.es}          
          \and Y.-J. Chen\inst{4}\email{asperchen@phy.ncu.edu.tw}
          }

   \institute{Centro de Astrobiolog\'{\i}a (CAB), INTA-CSIC, Ctra. de Ajalvir, km 4, Torrej\'on de Ardoz, 28850 Madrid, Spain
         \and Departamento de F\'{i}sica de la Tierra y Astrof\'{\i}sica, Facultad de Ciencias F\'{i}sicas, Universidad Complutense de Madrid, E-28040, Madrid, Spain
         \and Instituto de Qu\'{\i}mica F\'{\i}sica Blas Cabrera (CSIC), Serrano 119, 28006 Madrid, Spain
         \and Department of Physics, National Central University, Zhongli Dist., Taoyuan City 320317, Taiwan\\
             }

\date{}

\abstract
{Sulfur is the tenth most abundant element in space, yet it is strongly depleted in the gas phase of dense interstellar regions, suggesting sequestration in dust grain components that are difficult to detect.}
{We study whether UV irradiation and thermal processing of H$_2$S-containing interstellar-ice analogues can produce refractory sulfur species, and whether water-dominated ice enhances their formation.}
{We performed UV irradiation and warm-up experiments on H$_2$S-containing ice analogues with and without water, and the resulting refractory residues were analysed ex situ.}
{Refractory sulfur species, including the sulfur allotropes S$_6$--S$_8$, are formed during UV irradiation and thermal processing of H$_2$S-containing interstellar-ice analogues. Water enhances the formation of these sulfur allotropes by two orders of magnitude. A formation mechanism is proposed in which the ionic intermediates, identified in the reported experiments, stabilised by the water-ice matrix, promote chain growth and cyclisation, yielding predominantly S$_8$.}
{The combined detection in our residues of S$_\text{n}$ and polythionic species (H$_2$S$_\text{n}$O$_6$), also reported in meteorites and samples returned from asteroids, supports their formation in pre-cometary ice mantles covering dust grains. Because these species can incorporate multiple sulfur atoms per molecule, they constitute an efficient solid-phase reservoir for the ``missing sulfur'' in the dust of dense cloud environments where young stars are born. Refractory sulfur can therefore serve as a tracer of water-ice environments in space. We provide substantial evidence for the reaction pathway yielding sulfur deposits proposed in this work.}


\maketitle
\nolinenumbers

\section{Introduction}
\label{sect.introduction}

Sulfur is relatively abundant in the interstellar medium, with a cosmic abundance of \(1.23 \times 10^{-5}\) relative to hydrogen \citep{Snow1996}. In diffuse regions its abundance matches the cosmic value \citep{Sofia1994,Jenkins2009}, whereas in dense molecular clouds and protoplanetary discs the observed gas-phase sulfur represents only a fraction of the expected budget \citep{Anderson2013,Navarro2022}. This ``sulfur depletion problem'' suggests that sulfur is stored in dust grains or in carriers that are difficult to detect spectroscopically \citep{Mifsud2021,Taillard2025}. In this context, S$_8$ and other sulfur allotropes S$_\text{n}$ are plausible reservoirs because they sequester large numbers of sulfur atoms into stable species with intrinsically weak or elusive signatures.

Hydrogen sulfide (H$_2$S) is expected to be a key sulfur carrier in ice mantles on dust grains in cold regions. Although it has not been directly detected in the infrared \citep{Rocha2025}, models predict its presence in astrophysical ices \citep{Laas2019}. Laboratory simulations of pure H$_2$S ice and H$_2$S in an H$_2$O matrix \citep{Antonio2011,Cazaux2022,Hector2024,Mahjoub2024,Herath2025} show that products detected in the gas phase during warm-up following UV or electron irradiation include oxygenated sulfur species (e.g., SO$_2$, SO$_4^{2-}$, HSO$_3^{-}$, HSO$_4^{-}$), polysulfides H$_2$S$_\text{n}$ ($n = 2$--$11$), and sulfur allotropes S$_\text{n}$ ($n = 1$--$6$, 8). A key open issue is whether efficient formation of sulfur allotropes, particularly S$_8$, persists under water-dominated ice-mantle compositions where H$_2$S is a minor constituent.

While most species formed during the irradiation of astrophysical ice analogues desorb into the gas phase during heating, a fraction of the products can remain trapped in the solid at room temperature, forming a refractory residue (hereafter, the residue). The formation of refractory sulfur-bearing residues has been demonstrated by infrared spectroscopy in several laboratory studies. \citet{Guillermo2002} reported S$_6$--S$_8$ in residues from H$_2$O:CO:NH$_3$:H$_2$S ice mixtures; \citet{Antonio2014} found abundant S=O and C=S functionalities; and \citet{Pilling2015} observed bands compatible with sulfur oxoacids (H$_2$SO$_2$, H$_2$SO$_3$, $\mathrm{SO_4}^{2-}$) in residues from H$_2$O:CO$_2$:NH$_3$:SO$_2$ ice mixtures. The detection of some products both during warm-up (gas phase) and later in the solid residue suggests, as discussed by \citet{Hector2024}, that retention or desorption may depend on the non-covalent interactions established within the ice matrix. Nevertheless, the detailed molecular composition of sulfur residues and the potential role played by water in the formation of refractory sulfur species remain incompletely constrained.

Chromatographic analyses of samples from the carbonaceous asteroid (162173) Ryugu \citep{Takano2024,Aponte2023,Yoshimura2023} and of carbonaceous chondrites such as Murchison, Allende and Tagish Lake \citep{Randazzo2024,Naraoka2022} have revealed a broad suite of inorganic sulfur species, including sulfur allotropes (S$_6$, S$_7$ and S$_8$) and oxygen-bearing sulfur compounds. These include sulfur oxyanions ($\mathrm{SO_3}^{-}$, $\mathrm{S_2O_3}^{-}$, $\mathrm{HS_2O_3}^{-}$, $\mathrm{S_3O_3}^{-}$ and $\mathrm{S_4O_3}^{-}$), as well as oxoacids, dominated by $\mathrm{H_2SO_4}$, and polythionic species ($\mathrm{H_2S_nO_6}$ / $\mathrm{HS_nO_6}^{-}$, $n = 2$--$15$).

We conducted experiments using pure H$_2$S and binary mixtures of H$_2$O:H$_2$S (10:1) to reproduce the UV processing of pre-cometary ice analogues and evaluate how water controls the efficiency of sulfur allotrope production. Ice deposition and UV irradiation were performed simultaneously at $\sim$80~K to process relatively thick ice layers and enhance the formation of refractory products, followed by heating to room temperature. Infrared spectra of H$_2$O:H$_2$S (10:1) residues reveal the presence of oxygenated sulfur compounds, and GC--MS directly identifies the sulfur allotropes retained in the solid (S$_6$, S$_7$ and S$_8$). In addition, the GC--MS signal of S$_6$--S$_8$ is markedly enhanced in the water-rich experiment relative to pure H$_2$S, indicating that an H$_2$O-dominated matrix promotes the formation of sulfur allotropes, as discussed in Sect.~\ref{GC_MS}. Previous UV- and electron-irradiation studies of pure H$_2$S ice have shown the formation of polysulfides H$_2$S$_\text{n}$ \citep{Cazaux2022,Hector2024,Herath2025}, whereas our experiments indicate that embedding H$_2$S in a water-rich matrix submitted to UV irradiation favours sulfur allotropes S$_\text{n}$ as the main refractory components. Based on further analysis by electrospray ionisation mass spectrometry in negative ion mode (ESI--MS), we propose, in Sect.~\ref{mechanism}, a reaction scheme for the formation of S$_\text{n}$ and polythionic species (H$_2$S$_\text{n}$O$_6$) in pre-cometary ices. The proposed mechanism explains the coexistence of both types of sulfur compounds in meteorites and asteroid samples. The agreement between our inventory of sulfur species in the residues and that observed in primitive Solar System materials supports photochemical processing and thermal evolution of H$_2$S molecules embedded in water-rich ices as a viable pathway for forming refractory sulfur deposits that can be incorporated into planetary bodies.

\section{Methods}
\label{sect.experimental_setup}

\subsection{Ice irradiation experiments}

Experiments were performed using the MOSS (Mimicking Organic Synthesis in Space) setup at the Centro de Astrobiología (CAB, CSIC--INTA). MOSS is a high-vacuum chamber designed to reproduce the chemical evolution of pre-cometary ices; a detailed description is given in \citet{delBurgo2025}. The chamber operates at a base pressure $< 1 \times 10^{-7}$~mbar at room temperature. A KRS-5 (thallium bromoiodide) IR window mounted at the tip of a cold finger was used as the substrate and cooled to $\sim$80~K with liquid nitrogen. Gases and vapours were introduced through dosing lines directed towards the KRS-5 window. UV irradiation was provided by a deuterium lamp (X2D2 L9841, Hamamatsu), delivering a photon flux of $5 \times 10^{14}$~photons~cm$^{-2}$~s$^{-1}$ distributed across the 115--180~nm (10.78--6.89~eV) range at the sample position.

The starting materials were Milli-Q water (Millipore IQ-7000) and H$_2$S (Nippon Gases, 99.8\%). Ice mixtures were formed by simultaneous deposition and UV irradiation of pure H$_2$S or H$_2$O:H$_2$S (10:1) ices at a total pressure of $5 \times 10^{-5}$~mbar at 80~K for 6~h. After deposition and irradiation, the liquid-nitrogen supply was interrupted and the sample was warmed up to room temperature (290~K) at an average heating rate of 1.08~K~min$^{-1}$ between 84 and 234~K, and 0.081~K~min$^{-1}$ from 234~K to room temperature, leaving a refractory residue on the substrate. Once removed from the vacuum chamber, residues were characterised by Fourier Transform Infrared Spectroscopy (FTIR) in transmittance mode using a Bruker Vertex 70 spectrometer (resolution 1~cm$^{-1}$; 512 scans). Substrates were stored in sealed capsules under a nitrogen atmosphere until chromatographic analysis.

As the residue contains chemically distinct product families, complementary analytical techniques were required. GC--MS was applied to target mainly non-polar, neutral and semi-volatile sulfur species, whereas ESI--MS was used to characterise polar oxygenated sulfur compounds and other non-volatile sulfur species that are not amenable to GC analysis under our conditions.

\subsection{GC--MS analysis of ice residues}

Each residue was extracted from the substrate surface for 5~min using 1~mL of dichloromethane. The solvent was then evaporated to dryness at room temperature under a gentle N$_2$ stream, and reconstituted with 20~$\mu$L of dichloromethane. These concentrated solutions were injected (1~$\mu$L) in pulsed splitless mode (1~min, 30~psi) at 250~$^\circ$C into a Thermo Scientific Trace GC Ultra gas chromatograph coupled to a Thermo Scientific DSQ II mass spectrometer equipped with an electron ionisation (EI) source operating at 70~eV in full scan mode ($m/z$ 50--350). The interface and ion source temperatures were 275 and 200~$^\circ$C, respectively. Chromatographic separation was achieved on an RTX-5MS capillary column (30~m $\times$ 0.25~mm, 0.25~$\mu$m film thickness), using helium as the carrier gas at a constant flow of 1~mL~min$^{-1}$. The GC oven was programmed from 40~$^\circ$C (2~min) to 250~$^\circ$C at 10~$^\circ$C~min$^{-1}$, and held at 250~$^\circ$C for 30~min. Sulfur allotropes were identified based on their characteristic fragmentation patterns and isotopic distributions in the mass spectra.

\subsection{ESI--MS analysis of ice residues}
\label{Exp.esi_ms}

Electrospray ionisation mass spectrometry (ESI--MS) experiments were carried out on a Surveyor LC system coupled to an LCQ Deca mass spectrometer (Thermo Scientific). Residues were extracted by immersing the substrates in 1~mL of CH$_3$OH for 5~min. A 5~$\mu$L volume of the resulting solutions was introduced into the ESI--MS system by direct loop injection into a 25~$\mu$L~min$^{-1}$ flow of acetonitrile:water (75:25, v/v), without chromatographic separation. The ESI source was operated in the negative-ion mode with a spray voltage of 3~kV and a capillary temperature of 110~$^\circ$C. Mass spectra were recorded from $m/z$ 50 to 500.

\section{Results}
\label{Results}

\subsection{Infrared characterisation of residues from H$_2$S and H$_2$O:H$_2$S ices}
\label{IR_Results}

\begin{figure}
  \centering 
  \includegraphics[width=\linewidth]{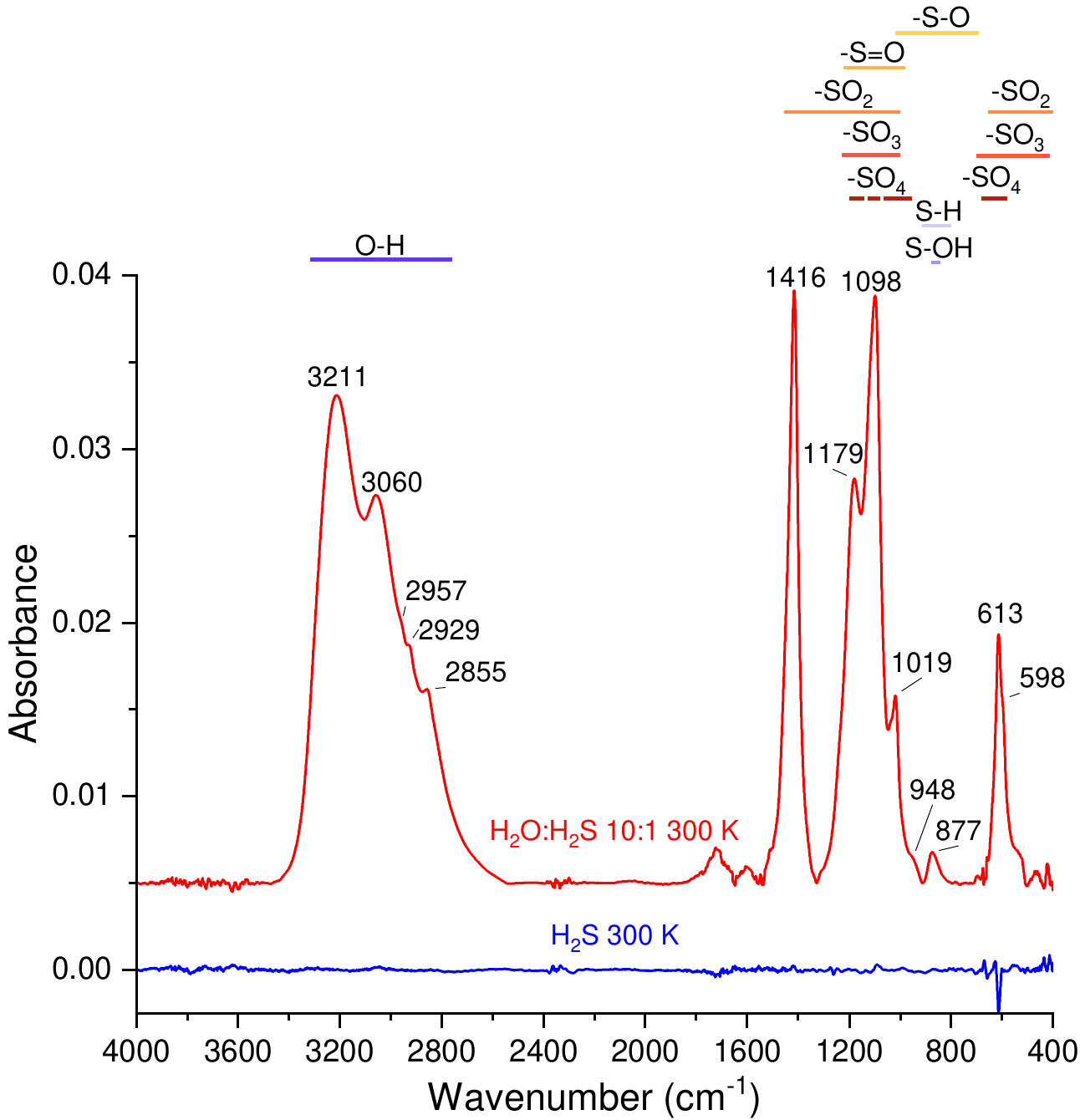}
  \caption{Infrared spectra of pure H$_2$S (blue) and H$_2$O:H$_2$S (10:1) (red) ice mixtures after UV irradiation followed by warm-up to room temperature, measured at 300~K. The residue from the H$_2$S ice does not show any IR absorption band. Functional groups assigned to bands in the H$_2$O:H$_2$S (10:1) residue are indicated at the top; vibrational bands and assignments are also listed in Table~\ref{Table.IR}, along with possible carrier species or chemical families producing those features.}
  \label{Fig.IR} 
\end{figure}

\begin{table*} 
\footnotesize
\centering
\caption[]{Proposed vibrational-mode assignments for the IR features present in the H$_2$O:H$_2$S (10:1) residue.}
\label{Table.IR}
\begin{tabular}{cccc}
\hline
\noalign{\smallskip}
$\tilde{\nu}$ (cm$^{-1}$) & Vibrational modes & Species & Ref.\\
\hline
\noalign{\smallskip}
3211 & O--H str. & $\mathrm{H_3O}^{+}$ & \cite{Strazzulla2007}\\
3060 & O--H str. & -- & --\\
2957 & O--H str. & $\mathrm{H_2S_{\text{n}}O_3}$ / $\mathrm{H_2SO_4}$ / $\mathrm{H_2S_{\text{n}}O_6}$ & \cite{Socrates2004}\\
2929 & O--H str. & $\mathrm{H_2S_{\text{n}}O_3}$ / $\mathrm{H_2SO_4}$ / $\mathrm{H_2S_{\text{n}}O_6}$ & \cite{Socrates2004}\\
2855 & O--H str. & $\mathrm{H_2S_{\text{n}}O_3}$ / $\mathrm{H_2SO_4}$ / $\mathrm{H_2S_{\text{n}}O_6}$ & \cite{Socrates2004}\\
1416 & SO$_2$ str. asym. & $\mathrm{H_2S_{\text{n}}O_3}$ / $\mathrm{H_2SO_4}$ / $\mathrm{H_2S_{\text{n}}O_6}$ & \cite{Socrates2004}\\
1179 & S=O str. / SO$_2$ str. sym. / SO$_3$ str. asym. & $\mathrm{H_2S_{\text{n}}O_3}$ / $\mathrm{H_2SO_4}$ / $\mathrm{H_2S_{\text{n}}O_6}$ & \cite{Socrates2004}; \cite{Guglielmero2025}\\
1098 & S=O str. / SO$_2$ str. sym. / SO$_3$ str. sym. / SO$_4$ str. & $\mathrm{H_2SO_4}$ & \cite{Socrates2004}; \cite{Pilling2015}\\
1019 & S--O str. asym. / S=O str. / SO$_3$ str. asym. & $\mathrm{H_2SO_4}$ & \cite{Socrates2004}\\
948  & S--O str. asym. & $\mathrm{H_2S_{\text{n}}O_3}$ / $\mathrm{H_2S_{\text{n}}O_6}$ & \cite{Socrates2004}; \cite{Guglielmero2025}\\
877  & S--O str. asym. / S--H def. / S--OH str. & $\mathrm{H_2S_{\text{n}}O_3}$ / $\mathrm{H_2SO_4}$ & \cite{Socrates2004}\\
613  & SO$_2$ def. / SO$_3$ bend. & $\mathrm{H_2S_{\text{n}}O_3}$ / $\mathrm{H_2SO_4}$ / $\mathrm{H_2S_{\text{n}}O_6}$ & \cite{Socrates2004}; \cite{Guglielmero2025}\\
598  & SO$_2$ str. / SO$_3$ bend. & $\mathrm{H_2SO_4}$ & \cite{Socrates2004}\\
\noalign{\smallskip}
\hline
\end{tabular}\\
{\footnotesize
\begin{flushleft}
str. = stretching; sym. = symmetric; asym. = asymmetric; def. = deformation; bend. = bending.\\
We have made assignments with the neutral species for simplicity, but the singly deprotonated and the doubled deprotonated anions are expected to be present and contribute to the IR features observed.
\end{flushleft}
}
\end{table*}

Figure~\ref{Fig.IR} shows the IR spectrum of the UV-irradiated pure H$_2$S experiment after warm-up (blue trace). No vibrational bands are detected, consistent with (i) extensive desorption of photoproducts before room temperature and (ii) intrinsically weak IR activity of S--S and S--H groups. In particular, S--S modes (540--400~cm$^{-1}$) are very weak because they induce only small dipole-moment changes \citep[see, e.g., S$_8$ IR band strength in][]{Taillard2025}; irradiation also favours long sulfur chains \citep{Cazaux2022,Hector2024,Herath2025}, reducing the relative abundance of terminal S--H bonds (2500--2400~cm$^{-1}$).

In contrast, the residue from the H$_2$O:H$_2$S (10:1) experiment shows intense absorptions (red trace; Fig.~\ref{Fig.IR}; Table~\ref{Table.IR}) indicative of oxygenated sulfur chemistry retained in the solid phase. Since IR spectroscopy does not allow us to determine the protonation state with precision, H$_2$SO$_4$, H$_2$S$_\text{n}$O$_3$ and H$_2$S$_\text{n}$O$_6$ are used here as generic terms encompassing H$_2$SO$_4$/HSO$_4^-$/SO$_4^{2-}$, H$_2$S$_\text{n}$O$_3$/HS$_\text{n}$O$_3^-$/S$_\text{n}$O$_3^{2-}$ and H$_2$S$_\text{n}$O$_6$/HS$_\text{n}$O$_6^-$/S$_\text{n}$O$_6^{2-}$, respectively. The spectrum displays a broad band between 3440--2560~cm$^{-1}$ associated with O--H stretching modes in different species; hydrated sulfonic acids (--S(=O)$_2$--OH) typically display very broad bands and may exhibit multiple maxima in this region \citep{Socrates2004}. The strong band at 1416~cm$^{-1}$ is consistent with the asymmetric stretching of the SO$_2$ group, which usually appears with very high intensity \citep{Socrates2004}. The bands at 1179, 948 and 613~cm$^{-1}$ agree with the frequencies calculated by \citet{Guglielmero2025} for the tetrathionate ion $\mathrm{S_4O_6}^{2-}$ (${}^{-}\mathrm{O_3S{-}S{-}S{-}SO_3^{-}}$), although other H$_2$S$_\text{n}$O$_6$ species ($n \geq 2$) would produce peaks at similar frequencies. The intense band at 1098~cm$^{-1}$ matches the characteristic stretching of ionic sulfate $\mathrm{SO_4}^{2-}$ \citep{Socrates2004}, which may result from deprotonation of $\mathrm{H_2SO_4}$, a compound reported by \citet{Pilling2015} in residues formed after broadband soft X-ray irradiation of H$_2$O:CO$_2$:NH$_3$:SO$_2$ ice mixtures. Overall, the spectral pattern is consistent with H$_2$SO$_4$, H$_2$S$_\text{n}$O$_3$ and H$_2$S$_\text{n}$O$_6$ species.

\subsection{Identification of sulfur allotropes by GC--MS}
\label{GC_MS}

\begin{figure*}
  \centering 
  \includegraphics[width=\linewidth]{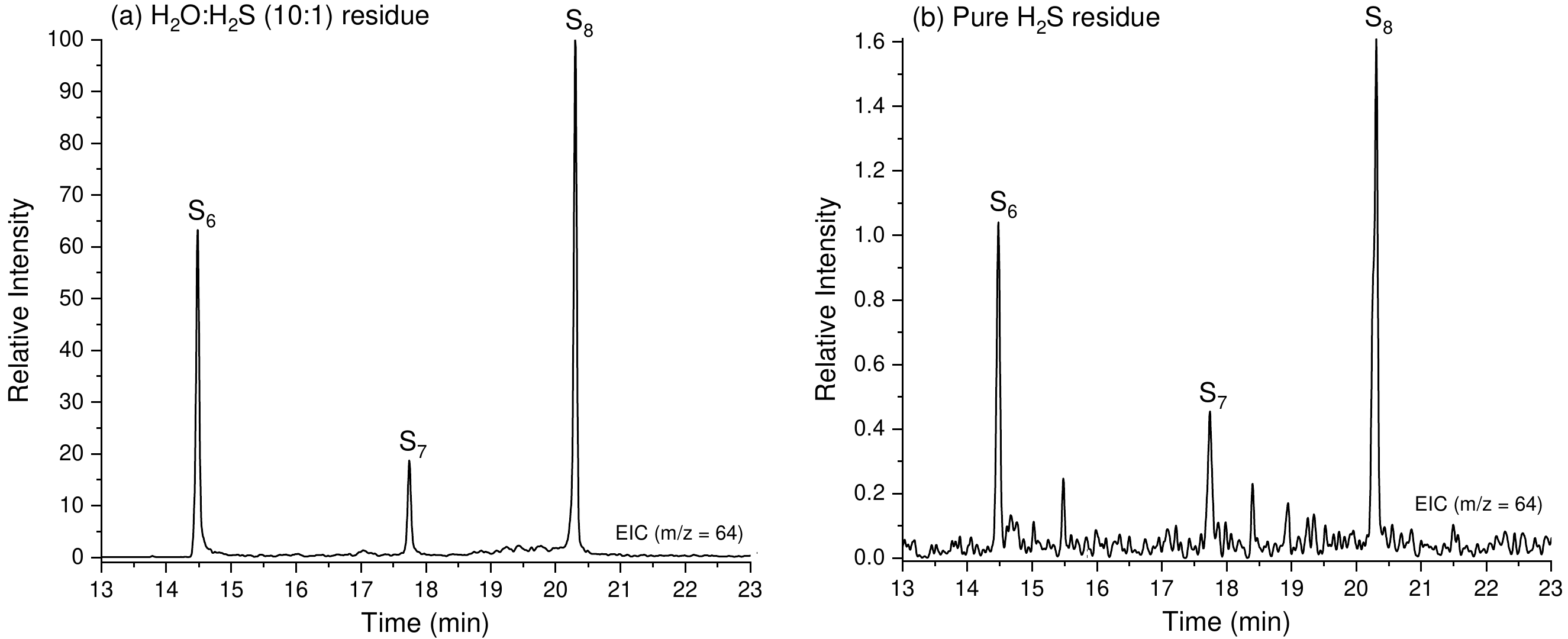}
  \caption{Extracted-ion chromatograms (EICs) for $m/z$~64 (S$_2^+$ fragment) obtained from GC--MS analyses of residues from (a) H$_2$O:H$_2$S (10:1) ice and (b) pure H$_2$S ice. In both samples, peaks assigned to the sulfur allotropes S$_6$, S$_7$ and S$_8$ are observed at 14.48, 17.74 and 20.30~min, respectively. For comparison, the y-axis is scaled in both panels to the most intense peak, S$_8$ in the H$_2$O:H$_2$S residue.}
  \label{Fig.GC} 
\end{figure*}

\begin{figure*}
  \centering 
  \includegraphics[width=\linewidth]{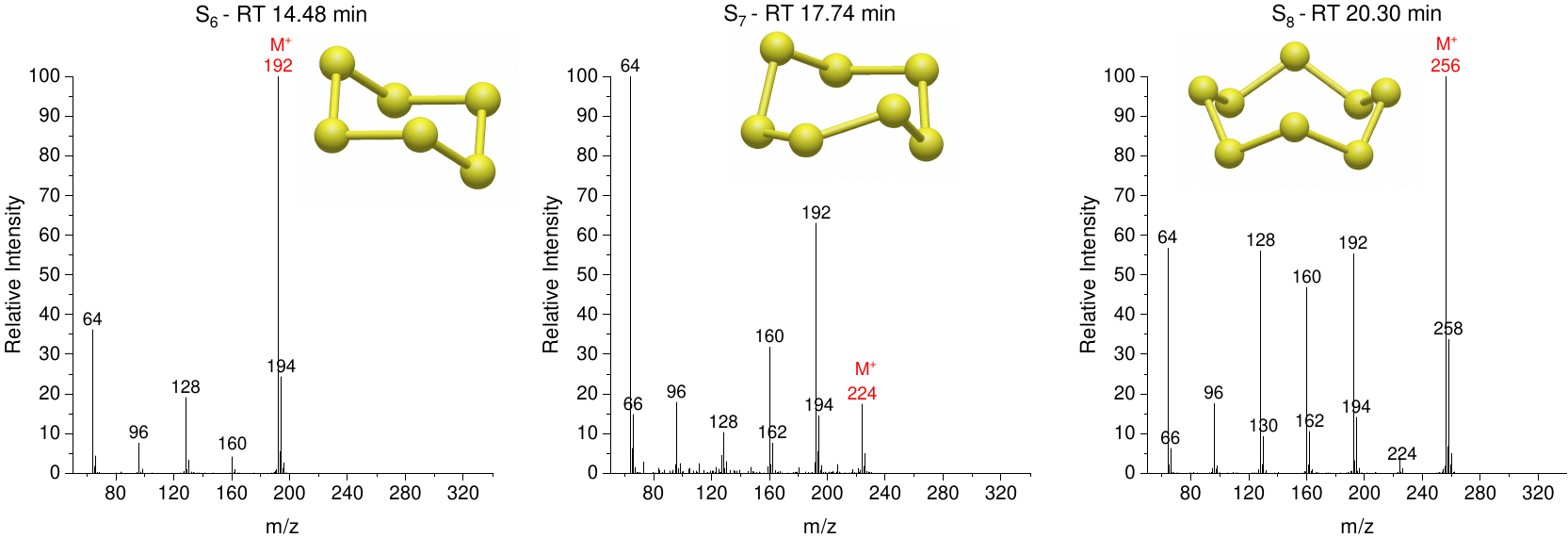}
  \caption{Mass spectra corresponding to compounds S$_6$, S$_7$ and S$_8$ detected by GC--MS in the residue of H$_2$O:H$_2$S (10:1). Each panel indicates the retention time and marks in red the molecular ion (M$^+$) corresponding to each species. For each compound, a structural representation of the assumed cyclic form, generated with Avogadro software, is included.}
  \label{Fig.MS} 
\end{figure*}

Figure~\ref{Fig.GC}a shows the extracted-ion chromatogram (EIC) of the H$_2$O:H$_2$S (10:1) ice mixture for $m/z$~64, corresponding to the S$_2^+$ fragment. It displays intense and well-defined peaks at retention times of 14.48, 17.74 and 20.30~min, whose mass spectra (Fig.~\ref{Fig.MS}) are assigned to the sulfur allotropes S$_6$, S$_7$ and S$_8$, respectively. Although GC--MS is not isomer-selective, we assume that these species adopt cyclic structures because ring isomers are the most commonly reported and thermodynamically favoured forms in this size range. Additionally, intramolecular ring closure should be much faster than intermolecular cyclisation to produce linear S$_8$. In particular, the crown conformer of S$_8$ has been independently identified by isomer-selective PI-ReToF-MS \citep{Herath2025}. In contrast, the chromatogram corresponding to the pure H$_2$S sample (Fig.~\ref{Fig.GC}b) displays very weak and noisy peaks at the same retention times. The mass spectra (not shown) associated with these features allow assignments to the sulfur allotropes S$_6$, S$_7$ and S$_8$, respectively.

Comparing the H$_2$S and H$_2$O:H$_2$S (10:1) experiments, the GC--MS peak areas of S$_6$--S$_8$ increase by factors of $\sim$50 (S$_6$), $\sim$40 (S$_7$), and $\sim$70 (S$_8$) in the water-rich ice relative to pure H$_2$S ice. Because the experiments were performed under high-vacuum conditions (base pressure $\sim 6 \times 10^{-6}$~mbar), the possible co-deposition of trace amounts of background H$_2$O during the experiment with pure H$_2$S cannot be entirely ruled out. Co-deposition of trace amounts of water molecules within the pure H$_2$S ice experiment would increase the formation yield of S$_\text{n}$. Consequently, the ratio of the S$_\text{n}$ formation between the H$_2$O:H$_2$S and the H$_2$S experiments will be decreased. An additional experiment with pure H$_2$S, carried out at a lower base pressure ($8 \times 10^{-8}$~mbar)---and therefore with a lower background deposition of water traces---showed an S$_8$ GC--MS signal about 1000 times lower than that obtained in the H$_2$O:H$_2$S (10:1) experiment. Partial sublimation of sulfur during the final warm-up cannot be excluded in either experiment. According to the low-pressure phase diagram reported by \citet{Ferreira2011}, at the pressures reached in our chamber, sulfur may enter the gas phase at temperatures around 285~K. Consequently, some S$_8$ may have been lost before recovery, although this loss cannot be quantified.

S$_8$ is the most abundant sulfur allotrope in both experiments, as shown in Fig.~\ref{Fig.GC}. Understanding the relative abundances of the sulfur allotropes S$_6$--S$_8$ and the yield contrast between ices with and without H$_2$O are the main objectives of this work. The mechanism explaining these observations will be discussed in Sect.~\ref{mechanism}.

\subsection{Identification of oxygenated sulfur species by ESI--MS}
\label{ESI_MS}

\begin{figure*}
  \centering 
  \includegraphics[width=\linewidth]{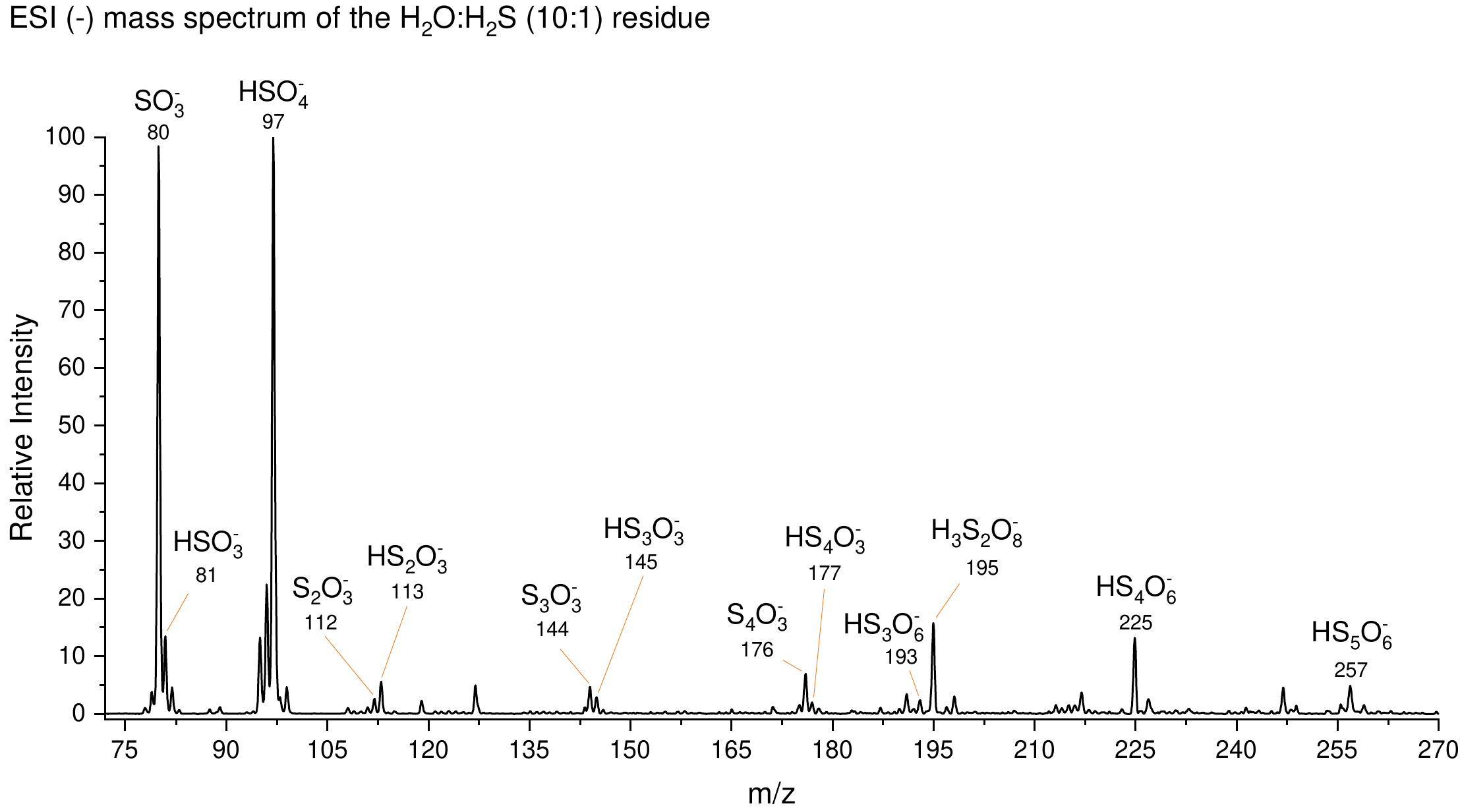}
  \caption{Direct injection ESI(-) full scan mass spectrum of the methanolic extract of the room-temperature residue from UV-processed H$_2$O:H$_2$S (10:1) ice. The spectrum shows ions assigned to the oxygenated sulfur-chain $\mathrm{S_{\text{n}}O_3}^{-}$, $\mathrm{HS_{\text{n}}O_3}^{-}$ and $\mathrm{HS_{\text{n}}O_6}^{-}$ families, together with signals from $\mathrm{HSO_4}^{-}$ and $\mathrm{H_3S_2O_8}^{-}$.}
  \label{Fig.ESI} 
\end{figure*}

To complement the IR assignments in Section \ref{IR_Results}, which suggest the presence of oxygenated sulfur species in the room temperature residue, a methanolic extract of the H$_2$O:H$_2$S (10:1) residue was analysed by negative-ion electrospray ionisation mass spectrometry (ESI($-$)--MS) using direct injection (i.e., without chromatographic separation). Figure \ref{Fig.ESI} shows the resulting full-scan ESI($-$) mass spectrum. The most intense sulfur-bearing ions are observed at $m/z$~80 and 97, assigned to SO$_3^{-}$ and HSO$_4^{-}$, respectively. Additional peaks are consistent with ions assigned to $\mathrm{HS_{\text{n}}O_3}^{-}$ ($\mathrm{HSO_3}^{-}$ to $\mathrm{HS_4O_3}^{-}$), $\mathrm{S_{\text{n}}O_3}^{-}$ ($\mathrm{S_2O_3}^{-}$ to $\mathrm{S_4O_3}^{-}$), and $\mathrm{HS_{\text{n}}O_6}^{-}$ ($\mathrm{HS_3O_6}^{-}$ to $\mathrm{HS_5O_6}^{-}$), together with $\mathrm{H_3S_2O_8}^{-}$. Although the ions detected by ESI($-$)--MS are deprotonated species, they may originate from compounds with different protonation states present in the residue (e.g., $\mathrm{S_{\text{n}}O_6}^{2-}$ / $\mathrm{HS_{\text{n}}O_6}^{-}$ / $\mathrm{H_2S_{\text{n}}O_6}$), depending on the local acid-base environment.

Assignments are supported by the observed $m/z$ values and by the agreement between experimental and theoretical isotopic patterns for the corresponding elemental composition. In particular, the presence of sulfur is corroborated by its characteristic isotopic signature, namely the detectable M+2 contribution arising from $^{34}$S at its natural abundance (4.47\%). Within the $\mathrm{HS_{\text{n}}O_6}^{-}$ series, $\mathrm{HS_4O_6}^{-}$ is the most intense ion, consistent with its comparatively higher stability \citep{Varga2007}. These results support the IR-based inference that the residue contains sulfate species (HSO$_4^{-}$) together with oxygenated sulfur-chain species consistent with the $\mathrm{HS_{\text{n}}O_3}^{-}$ and $\mathrm{HS_{\text{n}}O_6}^{-}$ families.

\section{Discussion}

\subsection{Formation mechanism of octoatomic sulfur (S$_8$)}
\label{mechanism}

\begin{figure*}
  \centering 
  \includegraphics[width=\linewidth]{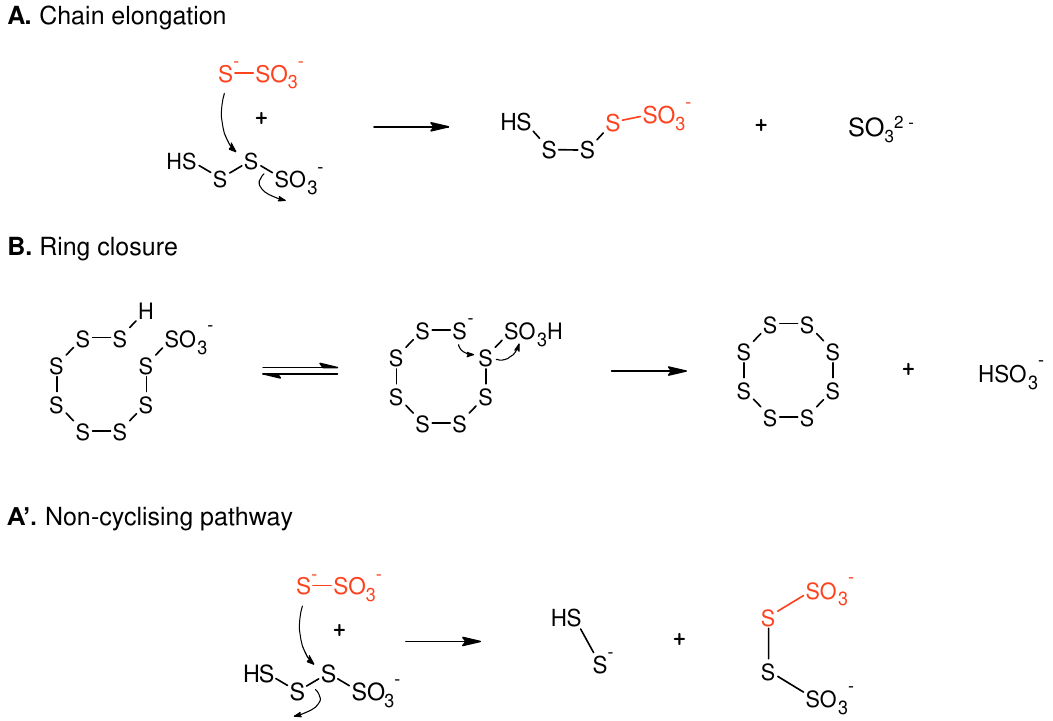}
  \caption{Proposed water-assisted mechanisms for sulfur chain growth and ring formation in H$_2$O:H$_2$S ice. 
(A) Chain elongation. Thiosulfate ($\mathrm{S_2O_3}^{2-}$, in red) attacks short sulfonated sulfur species (e.g., $\mathrm{HS_3}$--$\mathrm{SO_3}^{-}$), forming longer chains and releasing sulfite ($\mathrm{SO_3}^{2-}$). 
(B) Ring closure. Long sulfonated chains ($\mathrm{HS_9O_3}^{-}$) cyclise to yield elemental sulfur ($\mathrm{S_8}$) and bisulfite ($\mathrm{HSO_3}^{-}$). 
(A$^\prime$) Non-cyclising pathway. Attack at a different position yields polythionic species $\mathrm{S_{\text{n}}O_6}^{2-}$ / $\mathrm{HS_{\text{n}}O_6}^{-}$ / $\mathrm{H_2S_{\text{n}}O_6}$ rather than sulfur rings. A water-rich matrix promotes the generation and stabilisation of the oxygenated sulfur intermediates involved in these pathways.}
  \label{Fig.Mecanismo} 
\end{figure*}

The analytical results presented in Sect.~\ref{Results} reveal the molecular inventory retained in the room-temperature residue after UV processing of pure $\mathrm{H_2S}$ and H$_2$O:H$_2$S (10:1) ices at $\sim$80~K and subsequent warm-up. IR spectroscopy (Fig.~\ref{Fig.IR}; Table~\ref{Table.IR}) indicates oxygenated and sulfur-bearing functional groups (O--H, S--O, S=O, $\mathrm{SO_2}$, $\mathrm{SO_3}$, $\mathrm{SO_4}$, S--H and S--OH), whereas GC--MS (Figs.~\ref{Fig.GC} and \ref{Fig.MS}) identifies elemental sulfur allotropes $\mathrm{S_6}$--$\mathrm{S_8}$. Negative-ion ESI mass spectrometry (Fig.~\ref{Fig.ESI}) detects ions assigned to $\mathrm{SO_3}^{-}$ and $\mathrm{HSO_4}^{-}$, together with additional ions consistent with $\mathrm{HS_{\text{n}}O_3}^{-}$ ($n = 1$--$4$), $\mathrm{HS_{\text{n}}O_6}^{-}$ ($n = 3$--$5$), and $\mathrm{S_{\text{n}}O_3}^{-}$ ($n = 1$--$4$), as well as $\mathrm{H_3S_2O_8}^{-}$.

To interpret the presence of these products and, in particular, the dominant formation of $\mathrm{S_8}$, we propose a mechanism that combines radical chemistry induced by irradiation of $\mathrm{H_2S}$ and $\mathrm{H_2O}$ with subsequent nucleophilic displacement reactions between sulfurous species.

First, UV irradiation of the H$_2$O:H$_2$S ice mixture generates H$\cdot$ and HS$\cdot$ radicals from $\mathrm{H_2S}$, and H$\cdot$ and OH$\cdot$ radicals from $\mathrm{H_2O}$:

$$\refstepcounter{equation}
\mathrm{H_2S} \rightleftharpoons \mathrm{HS\cdot} + \mathrm{H\cdot}
\eqno(\theequation)$$

$$\refstepcounter{equation}
\mathrm{H_2O} \rightleftharpoons \mathrm{OH\cdot} + \mathrm{H\cdot}
\eqno(\theequation)$$

OH$\cdot$ radicals can recombine to form $\mathrm{H_2O_2}$, which can lead to HO$_2\cdot$ radical \citep{Gerakines1996}:

$$\refstepcounter{equation}
\mathrm{OH\cdot} + \mathrm{OH\cdot} \rightarrow \mathrm{H_2O_2} 
\eqno(\theequation)$$

$$\refstepcounter{equation}
\mathrm{H_2O_2} + h\nu \rightarrow \mathrm{HO_2\cdot} + \mathrm{H\cdot}
\eqno(\theequation)$$

Recombination between HS$\cdot$ and HO$_2\cdot$ leads to the formation of oxygenated sulfur species such as SO \citep{TSOLee1984}, which is subsequently oxidised to $\mathrm{SO_2}$ by reaction with oxygen atoms released from water photolysis or by oxygen transfer between SO molecules \citep{Moore2007}:

$$\refstepcounter{equation}
\mathrm{HS\cdot} + \mathrm{HO_2\cdot} \rightarrow \mathrm{SO + H_2O}
\eqno(\theequation)$$

$$\refstepcounter{equation}
\mathrm{SO} + \mathrm{O} \rightarrow \mathrm{SO_2} \quad \text{or} \quad \mathrm{SO} + \mathrm{SO} \rightarrow \mathrm{SO_2} + \mathrm{S}
\eqno(\theequation)$$

In the presence of water, $\mathrm{SO_2}$ becomes hydrated or protonated, yielding $\mathrm{HSO_3}^{-}$ \citep{Pilling2015}, a species that can also form by direct addition of OH radicals to $\mathrm{SO_2}$ \citep{Moore2007}:

$$\refstepcounter{equation}
\mathrm{SO_2} + \mathrm{H_2O} \rightleftharpoons \mathrm{HSO_3}^{-} + \mathrm{H}^{+}
\eqno(\theequation)$$

Previous detection of $\mathrm{SO_2}$ and $\mathrm{HSO_3}^{-}$ in UV-irradiated H$_2$O:H$_2$S ices \citep{Antonio2011}, together with our detection of $\mathrm{HSO_3}^{-}$ in the residue, supports this part of the reaction scheme. Finally, the deprotonation of $\mathrm{HSO_3}^{-}$ induced by irradiation or warming leads to the formation of $\mathrm{SO_3}^{2-}$ \citep{Pilling2015}:

$$\refstepcounter{equation}
\mathrm{HSO_3}^{-} \rightleftharpoons \mathrm{SO_3}^{2-} + \mathrm{H}^{+}
\eqno(\theequation)$$

As shown in Fig.~\ref{Fig.GC}, $\mathrm{S_8}$ formation is strongly enhanced in the presence of water. In particular, the intensity of the $\mathrm{S_8}$ signal in the pure $\mathrm{H_2S}$ residue is only 1.6\% of that observed in the H$_2$O:H$_2$S residue. As it will be described below, the generation of $\mathrm{SO_3}^{2-}$ (and hence sulfonated chain precursors, $\mathrm{HS_{\text{n}}O_3}^{-}$) is a key requirement for efficient $\mathrm{S_8}$ production. In the pure $\mathrm{H_2S}$ experiments, $\mathrm{SO_3}^{2-}$ can only arise at trace levels from residual O-bearing impurities in the system (e.g., $\mathrm{H_2O} / \mathrm{O_2}$), which limits the formation of $\mathrm{HS_{\text{n}}O_3}^{-}$ intermediates and results in much lower $\mathrm{S_8}$ yields than in H$_2$O:H$_2$S ices.

From this point, the synthesis of S$_8$ is expected to follow the classical pathways proposed by \citet{Davis1958} for the formation of elemental sulfur based on the chemistry of sulfonated species. The addition of HS$\cdot$ radicals or atomic sulfur to $\mathrm{SO_3}^{2-}$ leads to the formation of HS--$\mathrm{SO_3}^{-}$ species, in equilibrium with the thiosulfate ion $\mathrm{S_2O_3}^{2-}$. These HS--$\mathrm{SO_3}^{-}$ units constitute the fundamental building blocks for S--S chain growth. Theoretical calculations by \citet{Meyer1977} indicate that HS--$\mathrm{SO_3}^{-}$ is more stable than the $\mathrm{{}^{-}S{-}SO_3H}$ form, favouring its participation in subsequent reactions:

$$\refstepcounter{equation}
\mathrm{S\cdot} + \mathrm{SO_3}^{2-} \rightarrow \mathrm{S_2O_3}^{2-} \overset{\mathrm{H}^{+}}{\rightleftharpoons} \mathrm{HS\text{-}SO_3}^{-}
\eqno(\theequation)$$

$$\refstepcounter{equation}
\mathrm{HS\cdot} + \mathrm{SO_3}^{2-} \rightarrow \mathrm{HS\text{-}SO_3}^{-} \overset{\mathrm{H}^{+}}{\rightleftharpoons} \mathrm{S_2O_3}^{2-}
\eqno(\theequation)$$

In a water-rich matrix, these anionic intermediates are expected to be stabilised by solvation and counter-ions (e.g., $\mathrm{H_3O}^{+}$), which can facilitate subsequent reactions. Elongation of the S--S chain occurs through bimolecular nucleophilic substitution reactions, in which thiosulfate $\mathrm{S_2O_3}^{2-}$ acts as a nucleophile and preferentially attacks the sulfur atom adjacent to the --$\mathrm{SO_3}^{-}$ group, which is the most positively charged and sterically accessible site of the HS--$\mathrm{SO_3}^{-}$ chain \citep{Meyer1977}. This initial attack extends the S--S skeleton, forming HS--S--$\mathrm{SO_3}^{-}$ and expelling sulfite ($\mathrm{SO_3}^{2-}$), as illustrated in Fig.~\ref{Fig.Mecanismo}A. Successive attacks by $\mathrm{S_2O_3}^{2-}$ sequentially generate $\mathrm{HS_{\text{n}}O_3}^{-}$ species of increasing length:

$$\refstepcounter{equation}
\mathrm{HS_2O_3}^{-} + \mathrm{S_2O_3}^{2-} \rightarrow \mathrm{HS_3O_3}^{-} + \mathrm{SO_3}^{2-}
\eqno(\theequation)$$\\

or, in general:

$$\refstepcounter{equation}
\mathrm{HS_nO_3}^{-} + \mathrm{S_2O_3}^{2-} \rightarrow \mathrm{HS_{n+1}O_3}^{-} + \mathrm{SO_3}^{2-}
\eqno(\theequation)$$

When the chain reaches a sufficient length ($n \geq 7$), unimolecular cyclisation becomes competitive. \citet{Davis1958} specifically proposed this unimolecular closure for $\mathrm{HS_9O_3}^{-}$ yielding S$_8$ and for $\mathrm{HS_7O_3}^{-}$ yielding S$_6$. In this process (Fig.~\ref{Fig.Mecanismo}B), the terminal group --S--H transfers a proton to the --$\mathrm{SO_3}^{-}$ group, generating $\mathrm{HSO_3}^{-}$ as a good leaving group. The resulting deprotonated sulfur then intramolecularly attacks the sulfur atom adjacent to --SO$_3$H, leading to ring closure and the formation of an $\mathrm{S_{\text{n}-1}}$ sulfur allotrope. The charge-distribution calculations of \citet{Meyer1977} support the electrostatic plausibility of this intramolecular attack. Both steps could also occur in a concerted manner, through an unstable intermediate that readily evolves with release of a sulfur allotrope:

$$\refstepcounter{equation}
\mathrm{HS_nO_3}^{-} \rightarrow \mathrm{S_{n-1}} + \mathrm{HSO_3}^{-}, \quad n \ge 7 .
\eqno(\theequation)$$\\

or, specifically for $\mathrm{S_8}$:

$$\refstepcounter{equation}
\mathrm{HS_9O_3}^{-} \rightarrow \mathrm{S_8} + \mathrm{HSO_3}^{-}
\eqno(\theequation)$$

The formation of $\mathrm{S_8}$ is particularly favoured due to its higher thermodynamic and kinetic stability. Closure of smaller rings, such as $\mathrm{S_6}$ or $\mathrm{S_7}$, is less efficient because of their lower stability, favouring elongation of their precursor chains \citep{Davis1958}. Conversely, elongation beyond $\mathrm{HS_9O_3}^{-}$ and the subsequent formation of $\mathrm{S_{\text{n}}}$ ($n > 8$) are unlikely, as at this chain length unimolecular cyclisation leading to $\mathrm{S_8}$ will be a much faster process than further chain growth. This competition between elongation and cyclisation is consistent with S$_8$ being the most abundant sulfur allotrope detected in our GC--MS analyses (Fig.~\ref{Fig.GC}).

Support for this proposed mechanism is provided by the detection of several relevant products. The identification of sulfur allotropes by GC--MS is consistent with the proposed cyclisation step, although this ex situ analysis does not establish when ring closure occurred. Previous in situ experiments detected S$_8$ during TPD by PI-ReToF-MS in irradiated pure H$_2$S ice \citep{Herath2025}, and S$_{\text{x}}^{+}$ fragments consistent with S$_8$ desorption by TPD--QMS from processed pure H$_2$S and H$_2$O:H$_2$S ices \citep{Hector2024}. These observations show that S$_8$ can form before ex situ analysis. In the proposed mechanism, cyclisation may occur during irradiation, during the subsequent warm-up, or during both stages, when increasing molecular mobility can facilitate ring closure within the initially rigid matrix. In turn, detection of $\mathrm{SO_3}^{-}$ and $\mathrm{HSO_3}^{-}$ by ESI($-$)--MS is consistent with their recurrent release during chain elongation and closure. Such species can remain trapped in the solid matrix up to room temperature \citep{Pilling2015}, as intermolecular forces are stronger in this polar environment, accounting for their presence in the final residue. While SO$_3^{2-}$ is proposed as a key reactive intermediate, it is not directly observed in our ESI($-$)--MS spectra. In fact, it is indirectly detected as SO$_3^{-}$ and HSO$_3^{-}$. That can be explained as follows: SO$_3^{2-}$ is a relatively small species, carrying a high negative charge on it. This type of molecule undergoes rapid protonation (HSO$_3^{-}$), ion pairing, or both forming HSO$_3^{-}$ and SO$_3^{-}$ upon contact with the extraction solvent during ESI, favouring its detection as HSO$_3^{-}$ rather than as a doubly charged ion. Detection of short sulfonated intermediates $\mathrm{HS_{\text{n}}O_3}^{-}$ (e.g., $\mathrm{HSO_3}^{-}$, $\mathrm{HS_2O_3}^{-}$, $\mathrm{HS_3O_3}^{-}$ and $\mathrm{HS_4O_3}^{-}$) indicates that early reaction products can accumulate when formed in abundance and may not necessarily progress towards longer chains. Moreover, detection of the polythionic $\mathrm{HS_{\text{n}}O_6}^{-}$ family ($\mathrm{HS_3O_6}^{-}$, $\mathrm{HS_4O_6}^{-}$ and $\mathrm{HS_5O_6}^{-}$) reveals alternative pathways in which thiosulfate, $\mathrm{S_2O_3}^{2-}$, attacks internal positions of the $\mathrm{HS_{\text{n}}}$--$\mathrm{SO_3}^{-}$ chain (Fig.~\ref{Fig.Mecanismo}A$^\prime$). These attacks generate $\mathrm{SO_3}^{-}$--$\mathrm{S_{\text{n}}}$--$\mathrm{SO_3}^{-}$ species, which carry --$\mathrm{SO_3}^{-}$ groups at both ends and therefore lack an efficient leaving group such as $\mathrm{HSO_3}^{-}$, preventing cyclisation into sulfur rings and favouring the accumulation of these linear compounds.
 
The central result of this work is the role of water molecules in promoting $\mathrm{S_8}$ formation. As shown in Fig.~\ref{Fig.GC}, $\mathrm{S_8}$ is orders of magnitude more abundant in the water-rich ice than in pure $\mathrm{H_2S}$. This behaviour is explained by the need for $\mathrm{SO_3}^{2-}$ and $\mathrm{HSO_3}^{-}$ species in the formation of sulfur allotropes. In the absence of $\mathrm{H_2O}$, irradiation mainly leads to $\mathrm{H_2S_{\text{n}}}$ chains \citep{Cazaux2022,Hector2024,Herath2025}, whose terminal --SH groups are poor leaving groups and limit cyclisation, resulting in much lower yields of $\mathrm{S_8}$. The presence of water, by favouring the formation of $\mathrm{HS_{\text{n}}O_3}^{-}$ species, instead activates the elongation and cyclisation pathways responsible for the efficient production of $\mathrm{S_8}$ in H$_2$O:H$_2$S ices.

Since $\mathrm{SO_2}$ is an intermediate in the proposed route, pre-existing $\mathrm{SO_2}$ on ice mixtures such as $\mathrm{H_2S:SO_2}$ or $\mathrm{H_2S:SO_2:H_2O}$ ices could in principle modify the efficiency of sulfur-ring formation. An $\mathrm{H_2S:SO_2}$ ice may shorten the initial oxidation sequence leading to $\mathrm{HSO_3}^{-}$ and $\mathrm{SO_3}^{2-}$ precursors more efficiently, but efficient $\mathrm{S_8}$ formation is also expected to require reduced sulfur species from $\mathrm{H_2S}$, such as HS$\cdot$ or HS$\cdot$ radicals, together with stabilisation of ionic intermediates such as $\mathrm{HSO_3}^{-}$, $\mathrm{SO_3}^{2-}$, $\mathrm{S_2O_3}^{2-}$ and $\mathrm{HS_{\text{n}}O_3}^{-}$ being H$_2$O molecules the ones performing these roles in our experiments. Therefore, the relative efficiency of anhydrous $\mathrm{H_2S:SO_2}$ versus water-containing $\mathrm{H_2S:SO_2:H_2O}$ ices remains an open question that should be addressed in future experiments.

\subsection{Astrophysical implications}
\label{sect.implications}

Recent chromatographic analyses of samples from the carbonaceous asteroid (162173) Ryugu \citep{Takano2024,Aponte2023,Yoshimura2023}, together with studies of carbonaceous chondrites such as Murchison, Allende, and Tagish Lake \citep{Randazzo2024,Naraoka2022}, point to a rich inorganic sulfur inventory in primitive Solar System materials. Reported compounds include sulfur allotropes (S$_6$, S$_7$ and S$_8$) together with a range of oxygen-bearing sulfur species, encompassing sulfur oxyanions (e.g., SO$_3^{-}$, S$_2$O$_3^{-}$, HS$_2$O$_3^{-}$, S$_3$O$_3^{-}$ and S$_4$O$_3^{-}$) and strongly oxidised sulfur reservoirs dominated by H$_2$SO$_4$ and polythionic species ($\mathrm{H_2S_nO_6}$ / $\mathrm{HS_nO_6}^{-}$, $n = 2$--$15$).

The detection in our residues of species that match those observed in asteroids and meteorites suggests that a significant fraction of these compounds forms in water-rich interstellar ices containing H$_2$S via UV (and likely cosmic-ray) processing, and is then retained as refractory material through subsequent thermal evolution until incorporation into primitive Solar System bodies such as asteroids and comets.

The reaction scheme proposed in this work provides an explanation for the presence of elemental sulfur allotropes and polythionic species in extraterrestrial materials, where the initial abundance of sulfur-bearing species is relatively low. While H--S$_\text{n}$--SO$_3^-$ chains can cyclise to form S$_\text{n}$, an alternative pathway (Fig.~\ref{Fig.Mecanismo}A$^\prime$) yields polythionic species S$_\text{n}$O$_6^{2-}$ / HS$_\text{n}$O$_6^-$ / H$_2$S$_\text{n}$O$_6$ that are not expected to cyclise efficiently and therefore persist in the solid phase. This chemical bifurcation explains why intermediates HS$_\text{n}$O$_3^-$ are rarely reported in meteorites and, when observed, are expected to be dominated by short-chain members, potentially reflecting relatively limited processing of the parent object, whereas polythionic species can be preserved across a broader range of chain lengths.

The formation of sulfur allotropes shows a clear dependence on the presence of water. The enhanced production of S$_\text{n}$ species in H$_2$O:H$_2$S ices, compared to pure H$_2$S ices, implies that the detection of elemental sulfur in primitive bodies may serve as an indirect indicator of water-rich formation environments.

Our results also help reconcile previous UV- and electron-irradiation studies of H$_2$S-bearing ices \citep[e.g.][]{Cazaux2022,Hector2024,Mahjoub2024,Herath2025}. In particular, earlier works could not constrain a detailed pathway to S$_8$ formation and, in some cases, IR spectra showed little or no residue signal after warm-up. It is important to note that the IR features reported here correspond to minor products, as relatively thick ice samples are required for their detection. Nevertheless, these features provide key constraints for the mechanism proposed here for S$_\text{n}$ formation.     
Our GC--MS and ESI--MS analyses indicate that this mainly reflects sensitivity and yield limitations (i.e., the amount of refractory material available for analysis), rather than the absence of the underlying chemistry. Furthermore, we established the essential role played by water in the synthesis of S$_\text{n}$ allotropes.

These results support the idea that a substantial fraction of the missing sulfur in dense interstellar regions and protoplanetary environments may be sequestered in solid-phase reservoirs composed of sulfur allotropes (S$_\text{n}$) and polythionic species (H$_2$S$_\text{n}$O$_6$) formed in icy mantles, later incorporated into comets, asteroids, and meteorites. The detection of S$_8$ in Ryugu samples and in the Orgueil CI chondrite \citep{Aponte2023} is therefore consistent with the efficient formation and retention of S$_8$ in H$_2$O:H$_2$S ices, as inferred from our experiments.

\begin{acknowledgements}

This research was funded by projects PID2020-118974GB-C21 and PID2023-151513NB-C21 by the Spanish Ministry of Science and Innovation, and by the National Science and Technology Council, Taiwan, under grant no.~NSTC 114-2112-M-008-017-(Y.-J.C.). C.d.B.O. was supported by an INTA grant.

Author contributions. C.d.B.O., H.C., and G.M.M.C. conceived and designed the study and defined the experimental strategy. C.d.B.O. performed the ice experiments, residue extractions, and GC--MS analyses, analysed and interpreted the data, and drafted the manuscript. H.C. and G.M.M.C. guided the interpretation, supervised the analysis, and revised the manuscript. J.E.Q.-L. and R.L.-A. performed the ESI--MS and GC--MS analyses and contributed to the interpretation. Y.-J.C. contributed to the design and construction of the experimental setup and revised the manuscript. All authors discussed the results and approved the final manuscript.

\end{acknowledgements}

\section*{Data availability}
The data underlying this article will be shared on reasonable request to the corresponding author.

\bibliographystyle{aa}
\bibliography{bibliography} 

@ARTICLE{Snow1996,
       author = {{Snow}, Theodore P. and {Witt}, Adolf N.},
        title = "{Interstellar Depletions Updated: Where All the Atoms Went}",
      journal = {\apjl},
         year = 1996,
        month = sep,
       volume = {468},
        pages = {L65},
          doi = {10.1086/310225},
       adsurl = {https://ui.adsabs.harvard.edu/abs/1996ApJ...468L..65S}
}

@ARTICLE{Sofia1994,
       author = {{Sofia}, Ulysses J. and {Cardelli}, Jason A. and {Savage}, Blair D.},
        title = "{The Abundant Elements in Interstellar Dust}",
      journal = {\apj},
         year = 1994,
        month = aug,
       volume = {430},
        pages = {650},
          doi = {10.1086/174438},
       adsurl = {https://ui.adsabs.harvard.edu/abs/1994ApJ...430..650S}
}

@ARTICLE{Jenkins2009,
       author = {{Jenkins}, Edward B.},
        title = "{A Unified Representation of Gas-Phase Element Depletions in the Interstellar Medium}",
      journal = {\apj},
         year = 2009,
        month = aug,
       volume = {700},
       number = {2},
        pages = {1299-1348},
          doi = {10.1088/0004-637X/700/2/1299},
archivePrefix = {arXiv},
       eprint = {0905.3173},
 primaryClass = {astro-ph.GA},
       adsurl = {https://ui.adsabs.harvard.edu/abs/2009ApJ...700.1299J}
}

@ARTICLE{Anderson2013,
       author = {{Anderson}, Dana E. and {Bergin}, Edwin A. and {Maret}, S{\'e}bastien and {Wakelam}, Valentine},
        title = "{New Constraints on the Sulfur Reservoir in the Dense Interstellar Medium Provided by Spitzer Observations of S I in Shocked Gas}",
      journal = {\apj},
         year = 2013,
        month = dec,
       volume = {779},
       number = {2},
          eid = {141},
        pages = {141},
          doi = {10.1088/0004-637X/779/2/141},
       adsurl = {https://ui.adsabs.harvard.edu/abs/2013ApJ...779..141A}
}

@article{Navarro2022,
	author = {{Navarro-Almaida}, David},
	title = {The sulphur depletion problem in molecular clouds: The H2S case},
	DOI= "10.1051/epjconf/202226500032",
	url= "https://doi.org/10.1051/epjconf/202226500032",
	journal = {EPJ Web Conf.},
	year = 2022,
	volume = 265,
	pages = "00032",
}

@ARTICLE{Mifsud2021,
       author = {{Mifsud}, Duncan V. and {Ka{\r{A}}uchov{\'a}}, Zuzana and {Herczku}, P{\'e}ter and {Ioppolo}, Sergio and {Juh{\'a}sz}, Zolt{\'a}n and {Kov{\'a}cs}, S{\'a}ndor T.~S. and {Mason}, Nigel J. and {McCullough}, Robert W. and {Sulik}, B{\'e}la},
        title = "{Sulfur Ice Astrochemistry: A Review of Laboratory Studies}",
      journal = {\ssr},
         year = 2021,
        month = feb,
       volume = {217},
       number = {1},
          eid = {14},
        pages = {14},
          doi = {10.1007/s11214-021-00792-0},
archivePrefix = {arXiv},
       eprint = {2101.05001},
 primaryClass = {astro-ph.EP},
       adsurl = {https://ui.adsabs.harvard.edu/abs/2021SSRv..217...14M}
}

@article{Rocha2025,
  title={Ice inventory towards the protostar Ced 110 IRS4 observed with the James Webb Space Telescope-Results from the Early Release Science Ice Age program},
  author={Rocha, WRM and McClure, MK and Sturm, JA and Beck, TL and Smith, ZL and Dickinson, H and Sun, F and Egami, E and Boogert, ACA and Fraser, HJ and others},
  journal={A\&A},
  volume={693},
  pages={A288},
  year={2025},
  publisher={EDP Sciences}
}

@article{Laas2019,
	author = {{Laas}, Jacob C. and {Caselli}, Paola},
	title = {Modeling sulfur depletion in interstellar clouds⋆
},
	DOI= "10.1051/0004-6361/201834446",
	url= "https://doi.org/10.1051/0004-6361/201834446",
	journal = {A\&A},
	year = 2019,
	volume = 624,
	pages = "A108",
}

@ARTICLE{Antonio2011,
       author = {{Jim{\'e}nez-Escobar}, A. and {Mu{\~n}oz Caro}, G.~M.},
        title = "{Sulfur depletion in dense clouds and circumstellar regions. I. H$_{2}$S ice abundance and UV-photochemical reactions in the H$_{2}$O-matrix}",
      journal = {\aap},
         year = 2011,
        month = dec,
       volume = {536},
          eid = {A91},
        pages = {A91},
          doi = {10.1051/0004-6361/201014821},
archivePrefix = {arXiv},
       eprint = {1112.3240},
 primaryClass = {astro-ph.EP},
       adsurl = {https://ui.adsabs.harvard.edu/abs/2011A\&A...536A..91J}
}

@ARTICLE{Cazaux2022,
       author = {{Cazaux}, S. and {Carrascosa}, H. and {Mu{\~n}oz Caro}, G.~M. and {Caselli}, P. and {Fuente}, A. and {Navarro-Almaida}, D. and {Rivi{\'e}re-Marichalar}, P.},
        title = "{Photoprocessing of H$_{2}$S on dust grains. Building S chains in translucent clouds and comets}",
      journal = {\aap},
         year = 2022,
        month = jan,
       volume = {657},
          eid = {A100},
        pages = {A100},
          doi = {10.1051/0004-6361/202141861},
archivePrefix = {arXiv},
       eprint = {2110.04230},
 primaryClass = {astro-ph.GA},
       adsurl = {https://ui.adsabs.harvard.edu/abs/2022A\&A...657A.100C}
}

@article{Ferreira2011,
title = {The low-pressure phase diagram of sulfur},
journal = {J. Chem. Thermodyn.},
volume = {43},
number = {2},
pages = {95-104},
year = {2011},
issn = {0021-9614},
doi = {https://doi.org/10.1016/j.jct.2010.07.007},
url = {https://www.sciencedirect.com/science/article/pii/S0021961410002235},
author = {A.G.M. Ferreira and L.Q. Lobo}
}

@ARTICLE{Hector2024,
       author = {{Carrascosa}, H. and {Mu{\~n}oz Caro}, G.~M. and {Mart{\'\i}n-Dom{\'e}nech}, R. and {Cazaux}, S. and {Chen}, Y.-J. and {Fuente}, A.},
        title = "{Formation and desorption of sulphur chains (H$_{2}$S$_{x}$ and S$_{x}$) in cometary ice: effects of ice composition and temperature}",
      journal = {\mnras},
         year = 2024,
        month = sep,
       volume = {533},
       number = {1},
        pages = {967-978},
          doi = {10.1093/mnras/stae1768},
archivePrefix = {arXiv},
       eprint = {2407.14135},
 primaryClass = {astro-ph.EP},
       adsurl = {https://ui.adsabs.harvard.edu/abs/2024MNRAS.533..967C}
}

@article{Mahjoub2024,
doi = {10.3847/PSJ/ad7d86},
url = {https://doi.org/10.3847/PSJ/ad7d86},
year = {2024},
month = {oct},
publisher = {The American Astronomical Society},
volume = {5},
number = {10},
pages = {234},
author = {Mahjoub, Ahmed and Hodyss, Robert and Altwegg, Kathrin and Fayolle, Edith and Choukroun, Mathieu},
title = {Dual Sources of S2 Observed in Comet 67P: Insights from Comparing ROSINA Measurements and Laboratory Simulations},
journal = {PSJ},
}

@article{Herath2025,
author = {Herath, Ashanie and McAnally, Mason and Turner, Andrew and Wang, Jia and Marks, Joshua and Fortenberry, Ryan and Garcia-Alvarez, Jorge and Gozem, Samer and Kaiser, Ralf},
year = {2025},
month = {07},
pages = {},
title = {Missing interstellar sulfur in inventories of polysulfanes and molecular octasulfur crowns},
volume = {16},
journal = {Nat. Commun.},
doi = {10.1038/s41467-025-61259-2}
}

@phdthesis{Guillermo2002,
    author = {{Mu{\~n}oz Caro}, G.~M},
    title = {From photoprocessing of interstellar ice to
amino acids and other organics},
    school = {Leiden
Observatory},
    year = {2002}
}

@ARTICLE{Antonio2014,
       author = {{Jim{\'e}nez-Escobar}, A. and {Mu{\~n}oz Caro}, G.~M. and {Chen}, Y.-J.},
        title = "{Sulphur depletion in dense clouds and circumstellar regions. Organic products made from UV photoprocessing of realistic ice analogs containing H$_{2}$S}",
      journal = {\mnras},
         year = 2014,
        month = sep,
       volume = {443},
       number = {1},
        pages = {343-354},
          doi = {10.1093/mnras/stu1100},
       adsurl = {https://ui.adsabs.harvard.edu/abs/2014MNRAS.443..343J}
}

@article{Pilling2015,
doi = {10.1088/0004-637X/811/2/151},
url = {https://doi.org/10.1088/0004-637X/811/2/151},
year = {2015},
month = {sep},
publisher = {The American Astronomical Society},
volume = {811},
number = {2},
pages = {151},
author = {Pilling, S. and Bergantini, A.},
title = {THE EFFECT OF BROADBAND SOFT X-RAYS IN SO2-CONTAINING ICES: IMPLICATIONS ON THE PHOTOCHEMISTRY OF ICES TOWARD YOUNG STELLAR OBJECTS},
journal = {ApJ}
}

@article{Takano2024,
author = {Takano, Yoshinori and Naraoka, Hiroshi and Dworkin, Jason and Koga, Toshiki and Sasaki, Kazunori and Sato, Hajime and Oba, Yasuhiro and Ogawa, Nanako and Yoshimura, Toshihiro and Hamase, Kenji and Ohkouchi, Naohiko and Parker, Eric and Aponte, José and Glavin, Daniel and Furukawa, Yoshihiro and Aoki, Junken and Kano, Kuniyuki and Nomura, Shin-ichiro and Orthous-Daunay, Francois-Regis and Tachibana, Shogo},
year = {2024},
month = {07},
pages = {},
title = {Primordial aqueous alteration recorded in water-soluble organic molecules from the carbonaceous asteroid (162173) Ryugu},
volume = {15},
journal = {Nat. Commun.},
doi = {10.1038/s41467-024-49237-6}
}

@article{Aponte2023,
author = {Aponte, José and Dworkin, Jason and Glavin, Daniel and Elsila, Jamie and Parker, Eric and McLain, Hannah and Naraoka, Hiroshi and Okazaki, Ryuji and Takano, Yoshinori and Tachibana, Shogo and Dong, Guannan and Zeichner, Sarah and Eiler, John and Yurimoto, Hisayoshi and Nakamura, Tomoki and Yabuta, Hikaru and Terui, Fuyuto and Noguchi, Takaaki and Sakamoto, Kanako and Watanabe, Sei-ichiro},
year = {2023},
month = {02},
pages = {},
title = {PAHs, hydrocarbons, and dimethylsulfides in Asteroid Ryugu samples A0106 and C0107 and the Orgueil (CI1) meteorite},
volume = {75},
journal = {Earth Planets Space},
doi = {10.1186/s40623-022-01758-4}
}

@article{Yoshimura2023,
author = {Yoshimura, Toshihiro and Takano, Yoshinori and Naraoka, Hiroshi and Koga, Toshiki and Araoka, Daisuke and Ogawa, Nanako and Schmitt-Kopplin, Ph and Hertkorn, Norbert and Oba, Yasuhiro and Dworkin, Jason and Aponte, José and Yoshikawa, Takaaki and Tanaka, Satoru and Ohkouchi, Naohiko and Hashiguchi, Minako and McLain, Hannah and Parker, Eric and Sakai, Saburo and Yamaguchi, Mihoko and Tachibana, Shogo},
year = {2023},
month = {09},
pages = {5284},
title = {Chemical evolution of primordial salts and organic sulfur molecules in the asteroid 162173 Ryugu},
volume = {14},
journal = {Nat. Commun.},
doi = {10.1038/s41467-023-40871-0}
}

@article{Randazzo2024,
author = {Randazzo, N. and Hilts, R. W. and Holt, M. C. and Herd, C. D. K. and Reiz, B. and Whittal, R. M.},
title = {Detection and quantification of organosulfur species in the Tagish Lake Meteorite by highly sensitive LC-MS},
journal = {Meteorit. Planet. Sci.},
volume = {59},
number = {9},
pages = {2241-2259},
doi = {https://doi.org/10.1111/maps.14189},
url = {https://onlinelibrary.wiley.com/doi/abs/10.1111/maps.14189},
eprint = {https://onlinelibrary.wiley.com/doi/pdf/10.1111/maps.14189},
year = {2024}
}

@article{Naraoka2022,
author = {Naraoka, Hiroshi and Hashiguchi, Minako and Okazaki, Ryuji},
year = {2022},
month = {12},
pages = {},
title = {Soluble Sulfur-Bearing Organic Compounds in Carbonaceous Meteorites: Implications for Chemical Evolution in Primitive Asteroids},
volume = {7},
journal = {ACS Earth Space Chem.},
doi = {10.1021/acsearthspacechem.2c00157}
}

@article{delBurgo2025,
	author = {{del Burgo Olivares}, C. and {Carrascosa, H.} and {Muñoz Caro, G. M.} and {Muñoz-Arnanz, J.} and {González-Díaz, C.} and {Chen, Y.-J.}},
	title = {Synthesis of new hexamethylenetetramine-based products (HMT-R) in UV-irradiated pre-cometary ice analogues},
	DOI= "10.1051/0004-6361/202554613",
	url= "https://doi.org/10.1051/0004-6361/202554613",
	journal = {A\&A},
	year = 2025,
	volume = 698,
	pages = "A285",
}

@article{Moore2007,
title = {The radiolysis of SO2 and H2S in water ice: Implications for the icy jovian satellites},
journal = {Icarus},
volume = {189},
number = {2},
pages = {409-423},
year = {2007},
issn = {0019-1035},
doi = {https://doi.org/10.1016/j.icarus.2007.01.018},
url = {https://www.sciencedirect.com/science/article/pii/S0019103507000589},
author = {M.H. Moore and R.L. Hudson and R.W. Carlson},
}

@article{Strazzulla2007,
title = {Hydrate sulfuric acid after sulfur implantation in water ice},
journal = {Icarus},
volume = {192},
number = {2},
pages = {623-628},
year = {2007},
issn = {0019-1035},
doi = {https://doi.org/10.1016/j.icarus.2007.08.004},
url = {https://www.sciencedirect.com/science/article/pii/S0019103507003478},
author = {G. Strazzulla and G.A. Baratta and G. Leto and O. Gomis},
}

@book{Socrates2004,
  author    = {George Socrates},
  title     = {Infrared and Raman Characteristic Group Frequencies: Tables and Charts},
  edition   = {3},
  publisher = {John Wiley \& Sons},
  year      = {2004},
  isbn      = {9780470093078},
  pages     = {368},
}

@Article{Guglielmero2025,
AUTHOR = {Guglielmero, Luca and Becherini, Stefano and D’Andrea, Felicia and Guazzelli, Lorenzo and Pomelli, Christian Silvio and de Angelis, Alberto Renato and Maffeis, Ivan and Parker, Wallace O’Neil and Mezzetta, Andrea},
TITLE = {A Novel Approach for the Preparation of Tetrathionate Ionic Liquids and Study of Their Sulfur Dissolution Properties},
JOURNAL = {Int. J. Mol. Sci.},
VOLUME = {26},
YEAR = {2025},
NUMBER = {19},
ARTICLE-NUMBER = {9823},
URL = {https://www.mdpi.com/1422-0067/26/19/9823},
PubMedID = {41097087},
ISSN = {1422-0067},
DOI = {10.3390/ijms26199823}
}

@article{Gerakines1996,
author = {Gerakines, Perry and Schutte, W. and Ehrenfreund, Pascale},
year = {1996},
month = {07},
pages = {289-305},
title = {Ultraviolet processing of interstellar ice analogs: I. Pure ices},
volume = {312},
journal = {A\&A}
}

@article{TSOLee1984,
author = {Tso, Tai Ly and Lee, Edward K. C.},
title = {Formation of sulfuric acid and sulfur trioxide/water complex from photooxidation of hydrogen sulfide in solid oxygen at 15 K},
journal = {J. Phys. Chem.},
volume = {88},
number = {13},
pages = {2776-2781},
year = {1984},
doi = {10.1021/j150657a021},

URL = { 
    
        https://doi.org/10.1021/j150657a021
    
    

},
eprint = { 
    
        https://doi.org/10.1021/j150657a021
    
    

}

}

@article{Davis1958,
author = {Davis, Robert Earl},
title = {Displacement Reactions at the Sulfur Atom. I. An Interpretation of the Decomposition of Acidified Thiosulfate},
journal = {J. Am. Chem. Soc.},
volume = {80},
number = {14},
pages = {3565-3569},
year = {1958},
doi = {10.1021/ja01547a018},

URL = { 
    
        https://doi.org/10.1021/ja01547a018
    
    

},
eprint = { 
    
        https://doi.org/10.1021/ja01547a018
    
    

}

}

@article{Meyer1977,
author = {Meyer, B. and Peter, L. and Spitzer, K.},
title = {Trends in the charge distribution in sulfanes, sulfanesulfonic acids, sulfanedisulfonic acids, and sulfurous acid},
journal = {Inorg. Chem.},
volume = {16},
number = {1},
pages = {27-33},
year = {1977},
doi = {10.1021/ic50167a007},

URL = { 
    
        https://doi.org/10.1021/ic50167a007
    
    

},
eprint = { 
    
        https://doi.org/10.1021/ic50167a007
    
    

}

}

@misc{Taillard2025,
      title={Predicting the detectability of sulphur-bearing molecules in the solid phase with simulated spectra of JWST instruments}, 
      author={A. Taillard and R. Martín-Doménech and H. Carrascosa and J. A. Noble and G. M. Muñoz Caro and E. Dartois and D. Navarro-Almaida and B. Escribano and A. Sanchez-Monge and A. Fuente},
      year={2025},
      eprint={2502.09384},
      archivePrefix={arXiv},
      primaryClass={astro-ph.GA},
      url={https://arxiv.org/abs/2502.09384}, 
}

@article{Varga2007,
author = {Varga, Dénes and Horváth, Attila},
year = {2007},
month = {10},
pages = {7654-61},
title = {Kinetics and Mechanism of the Decomposition of Tetrathionate Ion in Alkaline Medium},
volume = {46},
journal = {Inorg. Chem.},
doi = {10.1021/ic700992u}
}
\end{document}